\documentstyle[titlepage]{article}
\newcommand{\R}{{\rm I\kern-2pt R}}

\newtheorem{proposition}{Proposition}[section]

\newtheorem{theorem}{Theorem}[section]
\newtheorem{remark}{Remark}[section]
\newtheorem{definition}{Definition}[section]

\newtheorem{algorithm}{Algorithm}[section]

\begin{document}
\begin{center}
{\bf On a Quaternionic Representation for $Sp(4, R)$\\} 
Yassmin R. Ansari $\&$ Viswanath Ramakrishna \\ 
Department of Mathematical Sciences
and Center for Signals, Systems and Communications\\
University of Texas at Dallas\\
P. O. Box 830688\\
Richardson, TX 75083 USA\\
email: vish@utdallas.edu 
\end{center}

\begin{abstract}
This work provides a quaternionic representation for
real symplectic matrices in dimension four, analogous
to that for the orthogonal group. It also provides a technique
to compute this representation from the entries of the matrix
being thus represented. As a byproduct it shows how one can
compute the polar decomposition of a $4\times 4$ symplectic
matrix without any recourse to spectral calculations. 

{\it PACS Numbers: 03.65.Fd, 02.10.Yn, 02.10.Hh}
\end{abstract}

\section{Introduction}
The most important groups in physics are, arguably, the rotation,
and the symplectic groups, \cite{squeezing,gilmore}. In particular,
the symplectic group is central to classical mechanics, \cite{liebermann},
to classical and quantum optics, \cite{squeezing}, quantum
mechanics and quantum information processing, \cite{goongi}.  
Therefore, having as many parametrizations
of these groups as possible is desirable. 

In dimension four, there is
a well known parametrization of the orthogonal group via a pair of
unit quaternions. This has innumerable applications in physics and
engineering, \cite{kuiper,pertii}.
For the symplectic group, $Sp(4, R)$,
there seems to be none. Note, by the ``symplectic group", we refer
to the real symplectic group, and not the similarly named group
$Sp(n)$, which preserves the standard inner product on $H^{n}$
(here $H$ stands for the quaternions). 
The latter is, of course, defined via quaternions, \cite{pertii}. 
It is the intention of this note to provide such a parametrization
for $Sp(4, R)$.
Since the symplectic group is 10 dimensional (in comparison to the
six dimensions of the $SO(4, R)$), the parametrization to be provided
is accordingly more intricate.

For motivating the balance of the paper (in particular, the contents of 
the introductory section), we assume temporarily
some familiarity with the algebra ismorphism between $H\otimes H$
(the tensor product of the quaternions, $H$, with itself) and the
algebra of $4\times 4$ real matrices, denoted $M_{4}(R)$,
\cite{pertii,kyf,nii,haconi,ni,niii,expistruc,expisufour}. See section
2 for more details. Using this algebra isomorphism, it would seem
that to obtain a quaternion representation of an $X\in Sp(4, R)$, a logical
approach would be to write out the conditions
imposed on the quaternion representation
of an $X\in M_{4}(R)$
by the relation
\[
 X^{T}J_{4}X = J_{4}
\]
where $J_{4}$ is the defining matrix
of the symplectic group (its quaternion representation is $1\otimes j$).
However, this produces an immensely complicated system of equations.

Indeed, if one were to pursue this approach for the orthogonal group,
i.e., write out the condition $X^{T}X = I$ in quaternion form, one
does not recover the pair of unit quaternions representation. This is
not surpirising since those orthogonal matrices with determinant equal
to $-1$ are not represented by decomposable tensors, i.e., by an element
of $H\otimes H$, of the special form $u\otimes v$, with $u,v\in H$. 
Only if one were
to insist that the quaternion representation of the $X$ one began with
was to be already given by a decomposable tensor, would one recover the fact
that an $X\in SO(4, R)$ is represented by a tensor $u\otimes v$, with
$u$ and $v$ unit quaternions. Note that the condition, ${\mbox det}(X) = 1$
for the group $SO(4, R)$ is
subsumed by this representation, since the determinant of a matrix
represented by $p\otimes q$, with $p,q\in H$, is 
\[
\mid\mid p\mid\mid^{4}
\mid\mid q\mid\mid^{4}
\]
as a calculation shows. 

However, it is not
immediately evident that a special orthogonal matrix must be represented
by a decomposable tensor. In our opinion, the simplest way to see this
is to use the fact that the exponential map from $so(4, R)$ (the Lie
algebra of $4\times 4$ anti-symmetric matrices) is onto $SO(4, R)$.
Since the quaternion reprsentation of anti-symmetric matrices is easy to
find (see section 2), a calculation, \cite{expistruc}, shows that indeed
matrices in $SO(4, R)$ must be reprsented by decomposable matrices.
Unfortunately, the analogous statement, viz., that the exponential
map from $sp(4, R)$ (the Lie algebra of Hamiltonian matrices) is
onto $Sp(4, R)$, is false. Even if it were to be true, the exponential
of Hamiltonian matrices do not typically admit a decomposable quaternion 
representation.
Thus one cannot make the analogous simplification for the 
symplectic group.
 
To obtain a different perspective on the complexity of
the question at hand, we note that
it is straightforward to get quaternionic representations of 
Hamiltonian matrices. 
This is because: i) Hamiltonian matrices
are expressible as $J_{2n}S$, with $S$ real symmetric, and ii) quaternionic 
representations of symmetric matrices, and 
$J_{4}$ are easy to find.
Now Hamiltonian matrices form the Lie algebra of symplectic matrices.
So it is reasonable to inquire if
a similar translation from orthogonal to symplectic
matrices is valid. However, this is
patently incorrect, i.e., it is not true that a symplectic
matrix can be represented as $J_{4}G$, with $G$ orthogonal.
There also seems to be no easy modification of this approach which will
do the job.

To understand, the approach taken in this work, we note that the
initial suggestion that one write out the quaternionic version of
the condition $X^{T}J_{4}X = J_{4}$ leads to something tractable, provided
one imposes the additional restriction that the 
original $X$ be also {\it real symmetric}. 
This is very useful since it is known that in the polar
decomposition $X=UP$, of a symplectic $X$, both the orthogonal factor
($U$) and the positive definite factor ($P$) are symplectic as well
(see \cite{gilmore,liebermann,structureddecompI}, for instance).
Since $P$ is positive definite it is symmetric, and this explains the
utility of starting with the assumption that $X$ is symmetric and symplectic. 

For a variety of
reasons, it is more convenient to obtain quaternionic representations
of symmetric and symplectic $X$ and then further refine these to 
obtain quaternionic characterizations of positive definite and symplectic
$X$. Therefore, we eschew one route to the latter, viz., using the
fact that a positive definite symplectic $X$ must be the exponential 
of a symmetric, Hamiltonian matrix (this exponentiation can be
performed in closed form, \cite{expistruc} - but it would not
yield the maximal benefit). Characterizing symmplectic, orthogonal matrices
is relatively easy. Combining the two one gets a quaternionic representation
of symplectic matrices. These are stated in Theorem (\ref{quaternionrepnI}) and
Theorem (\ref{quaternionrepnII}), with the latter being an improvement
of the former which takes into account the characeterization of positive
definite, symplectic matrices obtained 
in Theorem (\ref{positivedefinitesymplectic}) . 

Let us now consider the question of finding formulae for
these quaternions in terms of the entries of the given symplectic matrix
$X$. The first step is to find $P$. Now $P$ is not any arbitrary symplectic,
positive definite matrix. It is, as in the case of all polar decompositions,
the unique positive definite square root of $X^{T}X$. The traditional
route to finding $P$ involves 
diagonalizing $X^{T}X$ [see Remark (\ref{diagonalize})]. 
In this work we avoid this. Rather, we provide an explicit algorithm 
for finding
$P$ which requires only the solution 
of a very simple $2\times 2$ linear system of equations.
This is achieved by explicitly showing that there are only two 
real symmetric, symplectic matrices $H$, with positive trace, satisfying
$H^{2} = X^{T}X$. Since one of these must be $P$, we can easily find $P$.
For all this it is crucial to characterize symmetric (not necessarily
positive definite) symplectic matrices via quaternions, which explains
the route chosen in this work.
Of course, these results on symmetric, symplectic matrices are of independent
interest.
We note, however, that this work does indeed obtain a quaternionic
characterization
of positive-definite, symplectic matrices
[see Theorem (\ref{quaternionrepnII})]. In particular, a quaternionic
parametrization of two-mode squeezing operators is thereby obtained, since 
positive-definite symplectic matrices are precisely those matrices which
represent such operations, \cite{squeezing}.

There are, of course, other global factorizations of the symplectic
group, such as the Euler (Cartan) and Iwasawa decompositions,
\cite{squeezing,gilmore,iwasawai,iwasawaii,iwasawaiii}, 
which could have been used as starting points
for finding a quaternionic representation. But we found the polar
decomposition as the most useful, since the polar decomposition
of a matrix has innumerable applications, \cite{hhorni}. Thus, the fact
that the polar decomposition based representation 
in this work, also yields a simple, constructive procedure for
finding the 
polar decomposition itself as a byproduct played an important motivating role.

It is appropriate at this point to record some history of the linear algebraic
applications of the
isomorphism between $H\otimes H$ and $M_{4}(R)$.
This isomorphism is central to the theory of Clifford algebras, \cite{pertii}.
However, it is only relatively recently been put into use for
linear algebraic (especially numerical linear algebraic) purposes.
To the best of our knowledge the first instance seems to be the work
of \cite{kyf}, where it was used in the study of linear maps preserving the
Ky-Fan norm. Then in \cite{haconi}, this connection was used
to obtain the Schur canonical form explicitly for real $4\times 4$ 
skew-symmetric matrices. Next, is the work of \cite{nii,ni,niii}, wherein
this connection was put to innovative use for solving eigenproblems of
a variety of structured $4\times 4$ matrices (including symmetric matrices).
Finally, in \cite{expistruc,expisufour}, this isomorphism was used to
explicitly calculate the exponentials of a wide variety 
of $4\times 4$ matrices.

The balance of this manuscript is organized as follows. In the next section
some notation and preliminary results on symplectic matrices, positive
definite matrices and the algebra isomorphism between $H\otimes H$ and
$M_{4}(R)$ are collected. The next section contains all the results
in this work.  Proposition (\ref{symsymplectic}) provides the quaternion
characterization of symmetric and symplectic matrices, central to this
work. This is used in Theorem (\ref{quaternionrepnI}) to provide
a quaternion representation of $Sp(4, R)$, which is further refined
in Theorem (\ref{quaternionrepnII}). This refinement is based on a quaternionic
characterization of positive 
definite matrices in Theorem (\ref{positivedefinitesymplectic}).
The principal technical tool in the proof
of Theorem (\ref{positivedefinitesymplectic}) is that of squaring a symmetric,
symplectic matrix. This is also a key ingredient in the proof of   
Theorem (\ref{explicitsqrt}). This latter theorem provides an explicit
technique to calculate the polar decomposition of matrices in $Sp(4, R)$.
While the proof of this theorem is somewhat lengthy, the key point
is that it yields an algorithm for finding the positive definite factor
in the polar decomposition, which requires only the calculation of a few
inner products in $R^{3}$ and the solution of a $2\times 2$ linear system.
This is summarized in an algorithm. Next a different perspective is provided
on the paucity of symplectic, symmetric square roots of $X^{T}X$, for an
$X\in Sp(4, R)$. As a byproduct an explicit formula for the characteristic
polynomial of an $X\in Sp(4, R)$ is obtained. The final section offers
some conclusions.  
 \section{Notation and Preliminary Observations}
The following definitions, notations and results will 
be frequently met in this work:
\begin{itemize} 
\item $M_{4}(R)$ (also denoted $gl (4, R)$) is the algebra of real
$4\times 4$ matrices.
\item $J_{2n}$ is the $2n\times 2n$ matrix which, in block form,
is given by $J_{2n} = \left ( \begin{array}{cc}
0_{n} & I_{n}\\
-I_{n} & 0_{n}
\end{array}
\right )$. $Sp(2n, R)$ denotes the Lie group of symplectic matrices,
i.e., those $2n\times 2n$ matrices, satsifying $X^{T}J_{2n}X
= J_{2n}$. $sp(2n, R)$ is its Lie algebra, 
consisting of those real (resp. complex)
$2n\times 2n$ matrices which satisfy $X^{T}J_{2n} + J_{2n}X = 0$.
Such matrices are also called {\it Hamiltonian}. 

\item If $X\in Sp(2n, R)$, then $X^{-1} = -J_{2n}X^{T}J_{2n}$.
Furthermore, if $X\in Sp(2n, R)$ then $X^{T}$ is also in $Sp(2n, R)$.

\item Essential use of the following theorem will be made in this work
(see \cite{gilmore,liebermann,structureddecompI}):
\begin{proposition}
\label{polardecompi}
{\rm Let $X$ be a real symplectic matrix, and let $X = PQ$ be its
polar decomposition, with $P$ positive definite and $Q$
real orthogonal. Then $P$ and $Q$ are also real symplectic.}
\end{proposition}
See \cite{structureddecompI} for examples of other matrix groups for
which an analogous statement holds. It is worth recalling here that
$P$ is the unique positive definite square root of the positive definite
matrix $X^{T}X.$
\item For a polynomial $P(x) = \sum_{i=0}^{n}a_{i}x^{i}$, of degree
at most $n$, its reverse is the polynomial $P_{{\mbox rev}}(x)
= \sum_{i=0}^{n}a_{n-i}x^{i}$. In this work we will use the fact that the
characteristic polynomial of a symplectic matrix {\it equals its reverse} 
[see \cite{structureddecompI}, for instance]. This follows from the 
well-known fact that for any invertible matrix $m\times m$ $A$, with 
characteristic polynomial 
$p(x)$, the characteristic polynomial of $A^{-1}$ is
the polynomial $(-1)^{m} ({\mbox det}(A))^{m}x^{m}p(\frac{1}{x})$.
Now for any polynomial, $x^{m}p(\frac{1}{x})$ is precisely its reverse.
Further for symplectic $A$, $m$ is even and ${\mbox det}(A)$ is $1$.
Finally, a symplectic $A$ is similar to its transpose. Hence,
the characteristic polynomial of $A$ is the same as that of its inverse,
whence it equals its reverse.   
\end{itemize}
We next collect some definitions and results on real positive 
definite matrices. All details, together with extensions to the
complex positive semidefinite case, may be found in \cite{hhorni}.
\begin{itemize}
\item 
\begin{definition}
\label{sqrts}
{\rm Let $Y$ be a real positive definite matrix. A real square matrix
$Z$ satisfying $Y = Z^{T}Z$ is said to said to be a square root of $Y$.}
\end{definition}
See \cite{hhorni} for details (extending to the case of complex positive
semidefinite matrices).
Square roots of positive definite matrices are not unique. However,
if $Z_{1}$ is a square root of $Y$ then $Z_{2}$ is also a square root
of $Y$ iff there exists a real orthogonal matrix $U$ such that $Z_{2}
= Z_{1}U$.
\item Let $Y$ be a real positive definite matrix. Then there exist
real symmetric matrices $H$ such that $Y = H^{2}$. Clearly any such
$H$ is a square root in the sense of Definition (\ref{sqrts}).

\item Let $Y$ be a real positive definite matrix. Then there exists a
unique real positive definite matrix $P$ with $Y = P^{2}$. Clearly
this $P$ is an example of a real symmetric matrix whose square equals $Y$. 

\end{itemize}

\noindent Next relevant definitions and results regarding
quaternions and their connection to real matrices will be presented.
Throughout $H$ will be denote the skew-field (
the division algebra) of the {\it quaternions},
while $P$ stands for the
{\it purely imaginary} quaternions, tacitly identified with $R^{3}$.

\noindent {\bf $H\otimes H$ and $M_{4}(R)$}: The algebra isomorphism
between $H\otimes H$ and $gl(4, R)$, which is central to this work is
the following:

\begin{itemize}
\item Associate to each product
tensor $p\otimes q\in H\otimes H$,
the matrix, $M_{p\otimes q}$, of the map which sends $x\in H$
to $px\bar{q}$, identifying $R^{4}$ with $H$ via the basis $\{1,i,j,k\}$.
Thus, if $p = p_{0} + p_{1}i + p_{2}j + p_{3}k; 
q = q_{0} + q_{1}i + q_{2}j + q_{3}k$, then
\[
M_{p\otimes q} = \left ( \begin{array}{cccc}
u_{0} & v_{0} & w_{0} & z_{0}\\
u_{1} & v_{1} & w_{1} & z_{1}\\
u_{2} & v_{2} & w_{2} & z_{2}\\
u_{3} & v_{3} & w_{3} & z_{3} 
\end{array}
\right )
\]
with
\begin{eqnarray*} 
p\bar{q}&=&u_{0} + u_{1}i + u_{2}j + u_{3}k\\
pi\bar{q}&=&v_{0} + v_{1}i + v_{2}j + v_{3}k\\
pj\bar{q}&=&w_{0} + w_{1}i + w_{2}j + w_{3}k\\
pk\bar{q}&=&z_{0} + z_{1}i + z_{2}j + z_{3}k
\end{eqnarray*}

Here, $\bar{q} = q_{0} - q_{1}i - q_{2}j - q_{3}k$

\item Extend this to the full tensor product by linearity,
e.g., the matrix associated to $2(p_{1}\otimes q_{1}) -
9(p_{2}\otimes q_{2})$ is the matrix $2M_{p_{1}\otimes q_{1}}
- 9 M_{p_{2}\otimes q_{2}}$. This yields an algebra isomorphism
between $H\otimes H$ and $M_{4}( R)$.
\item In particular, a basis for
$gl(4, R)$ is provided by the sixteen matrices $M_{e_{x}\otimes e_{y}}$
as $e_{x}, e_{y}$ run through $1, i, j, k$. Of these only $M_{1\otimes 1}$
is not traceless. 
In particular, $J_{4} = M_{1\otimes j}$ belongs to
this basis.
\item Define conjugation in $H\otimes H$ by first defining
the conjugate 
of a decomposable tensor $a\otimes b$ as $\bar{a}\otimes \bar{b}$, and then   
extending this to all
of $H\otimes H$ by linearity. Furthermore $M_{\bar{a}\otimes\bar{b}}
= (M_{a\otimes b})^{T}$.

\item Thus, the most general element of $M_{4}(R)$ admits the quaternion
representation $a1\otimes 1 + p\otimes i + q\otimes j + r\otimes k
+ s\otimes 1 + 1\otimes t$, with $a\in R$ and $p,q,r,s,t \in P$.
The summand $a1\otimes 1 + p\otimes i + q\otimes j + r\otimes k$
is the symmetric part of the  matrix, while the summand
$ s\otimes 1 + 1\otimes t$ is the skew-symmetric part of the matrix.
Expressions for $a,p,q,r,s,t$ (which are linear in the entries of
the matrix being represented) are easy to find, \cite{ni}. Finally
$4a$ is the trace of the matrix.  
\end{itemize}

\section{Quaternion Representations of $Sp(4, R)$}

In this section the main results of this work are presented.

To develop a quaternionic representation of an $X\in Sp(4, R)$, let us
invoke Theorem (\ref{polardecompi}). Per that result it suffices
to obtain quaternionic representations of positive definite, symplectic
matrices and symplectic, orthogonal matrices. 

Let $X = UP$ be the polar decomposition of $X$.
Since $U$ is symplectic and orthogonal it must, in fact, be 
special orthogonal.
Obtaining the quaternionic representation of such a matrix is easy.
It is given by $q = u\otimes v$, with $u, v$ unit quaternions
(corresponding to special orthogonality)  with
the further restriction that $vj = jv$ (corresponding to symplecticity). 
This imposes no further restriction on $u$, but forces $v$ to be of the form
$v = v_{0} + v_{2}j$, with $v_{0}^{2} + v_{2}^{2} = 1$.

To obtain a quaternionic representation of $P$, it is convenient
(for the reasons mentioned in Section 1) to obtain a quaternionic 
characterization of symmetric, symplectic $X$'s. This is achieved in:

\begin{proposition}
\label{symsymplectic}
{\rm Let $X$ be a $4\times 4$ symplectic matrix which is also symmetric.
Then it admits the quaternion representation
$X = a1\otimes 1 + p\otimes i + q\otimes j + r\otimes k$, with
$aq =r\times p$, $p.q = 0 = r.q$, and $a$ satisfying the constraint
$a^{2}-p.p + q.q - r.r = 1$. If $a$, which is $\frac{1}{4}$'th the trace 
of $X$, is not zero then $X$ is symplectic iff 
$aq = r\times p$ and $a^{2}-p.p + q.q - r.r = 1$.
}
\end{proposition}

{\it Proof:} The proof proceeds by equating the quaternion
expansion of $X^{T}J_{4}X = XJ_{4}X$, viz.,
$(a\ 1\otimes 1 + p\otimes i + q\otimes j + r\otimes k)(1\otimes j)
(a\ 1\otimes 1 + p\otimes i + q\otimes j + r\otimes k)$ with $1\otimes j$.

This calculation is facilitated by the observation that for any matrix
$X$ (not necessarily symplectic or symmetric), the matrix $X^{T}J_{2n}X$ 
is always skew-symmetric. Therefore, in the expansion 
$(a\ 1\otimes 1 + p\otimes i + q\otimes j + r\otimes k)(1\otimes j)
(a\ 1\otimes 1 + p\otimes i + q\otimes j + r\otimes k)$ one needs to
inspect only those terms of the form $s\otimes 1$ and $1\otimes t$, with
$s,t \in P$  
(i.e., those corresponding to the skew-symmetric terms) and equate all
but one of them to zero. The non-zero term is, of course, the $1\otimes j$ 
term and this is equated to $1$. Specifically, expansion of
$X^{t}J_{4}X$ is 
\[
(a^{2} - p.p + q.q - r.r)(1\otimes j) + (2r\times p - 2aq)\otimes 1
+ (2p.q)1\otimes i + (2r.q)1\otimes k
\]
Note that the vanishing
of the second term yields the condition $aq = r\times p$. Thus, if $a\neq 0$,
then either $q=0$ or $q = \frac{r\times p}{a}$ and this,
of course, ensures the conditions $p.q = 0 = r.q$. 
This finishes the proof.

\vspace*{9mm}

\noindent Now note that for the situation at hand, where 
$X$ is not merely symmetric but,
in fact, positive definite we necessarily have $a\neq 0$.
This yields a preliminary result about the quaternion representation
of symplectic matrices, which will be refined later in this section: 

\begin{theorem}
\label{quaternionrepnI}
{\rm Let $X$ be an element of $Sp(4,R)$. Then there exist $a, v_{0},
v_{2}\in R$,
$p, q, r\in P$, and a unit quaternion $u$ satisfying the
constraints $a^{2}-p.p + q.q - r.r = 1$, $q= \frac{r\times p}{a}$
and $v_{o}^{2} + v_{2}^{2}
= 1$ such that $X$ admits the following quaternion representation
$X = [u\otimes (v_{0} + v_{2}j)] 
[ a\ 1\otimes 1 + p\otimes i + q\otimes j +
r\otimes k]$.}
\end{theorem}

\noindent Note that this result 
gives the correct dimension count for $Sp(4, R)$.
$u$ is determined by $3$ real quantities and $v_{0}, v_{2}$ together
yield one additional parameter. The non-compact portion of $Sp(4, R)$ 
is determined by the two vector quantities $p, r$, and the scalar $a$, with
one restriction, thereby yielding $6$ real parameters.
Put together we get the dimension count to be $10$.

\vspace*{9mm}

\noindent We now turn to the issue of 
computing these quaternions from the entries
of $X$. Assuming the symmetric portion of the representation is available,
the calculation of the factor $[u\otimes (v_{0} + v_{2}j)]$ is 
amenable to the any technique which will yield the quaternion
representation of a matrix in $SO(4, R)$. This ought to be folklore,
but surprisingly the only explicit record of this that we were able
to find is in \cite{FourportI}. 

Next, consider computing the symmetric, $[ a\ 1\otimes 1 + p\otimes i 
+ \frac{r\times p}{a}\otimes j +
r\otimes k]$ factor. This is, of course, not just any symmetric matrix. 
It is the unique positive definite square root of $X^{T}X$.
To that end, one could, of course, resort to diagonalizing the symmetric
matrix $X^{T}X$. 
Instead, as we will see below that
there is an even more explicit method for finding the symmetric factor,
which reveals some facts which are of interest in their own right.

Specifically, recall from Section 2, that any positive 
definite matrix is the square
of a real symmetric matrix. The real symmetric matrix is not unique,
but one of these is the unique positive definite square root of the
positive definite matrix in question. Now, when the positive matrix
is $X^{T}X, X\in Sp(4, R)$, we know that the unique positive definite
square root is also symplectic. Furthermore, being
positive definite its trace is positive. So, instead of looking at all possible
real symmetric matrices whose square is $X^{T}X$, one needs to inspect
only those which are symplectic and have positive 
trace, in addition. One then finds the pleasant
conclusion that there are only few such candidates.

\begin{theorem}
\label{explicitsqrt}
{\rm Let $X\in Sp(4, R)$. Then there are atmost two (and, at least one) 
matrices $H$ which satisfy i) $H^{2} = X^{T}X$;
ii) $H$ is real symmetric and symplectic; and iii) ${\mbox Trace}\ (H)
> 0$. One of these is precisely the unique positive definite
square root of $X^{T}X$, and thus the positive definite factor in the polar
decomposition of $X$. Furthermore, $H$ can be found explicitly via
the solution of a simple linear systems.}
\end{theorem}

{\it Proof:} Let $X\in Sp(4, R)$. Then so is $X^{T}$ and hence $X^{T}X\in
Sp(4, R)$. Let $b1\otimes 1 + c\otimes i +  d\otimes j + e\otimes k$
be its quaternion representation. Note, as $X^{T}X$ is positive definite,
$b > 0$, while $c,d,e$ are pure quaternions. 
Let $H$ be a real symmetric, symplectic matrix with non-zero trace
satisfying $H^{2} = X^{T}X$. Suppose $H = a1\otimes 1 + p\otimes i
+ q\otimes j + r\otimes k$.
Then equating $H^{2}$ to $X^{T}X$ yields the system of equations: 

\begin{eqnarray}
\label{first}
b &=& a^{2} + p.p + q.q + r.r\\ \nonumber
c &=&  2ap + 2q\times r\\ \nonumber
d &=& 2aq + 2r\times p\\ \nonumber
e &=& 2ar + 2p\times q 
\end{eqnarray}
    
In addition, as $H$ is symplectic, this system is augmented by the
conditions i) $a >0$; ii) $q=\frac{r\times p}{a}$; iii) $1 = a^{2}
+ q.q - p.p - r.r$.

The second equation in the system Equation (\ref{first}), 
together with $2aq = r\times p$, yields
$q = \frac{d}{4a}$. So, if we can find
$a$ (note, we know that a solution with $a > 0$ exists, in
principle) then we obtain $q$. 

At this point it is convenient to divide the argument into two cases.
The case $q\neq 0$ and the case $q=0$. Note (as $a\neq 0$) these are equivalent
to the cases $d\neq 0$ and $d = 0$ respectively. 

\vspace*{5mm}

\noindent {\it The case $d \neq 0$:}
To find $a$, it is first noted that $H$ is symplectic, we have
\[
1 = a^{2} + q.q -p.p -r.r
\]
and hence 
$\frac{b+1}{2} = a^{2} + q.q = a^{2} + \frac{d.d}{16a^{2}}$.
This yields a quadratic for $a^{2}$, viz.,

\begin{equation}
\label{quadratic}
a^{4} - \frac{b+1}{2}a^{2} + \frac{d.d}{16} = 0 
\end{equation}

Note that the discriminant of the above equation is
$\frac{(b+1)^{2}}{4} - \frac{d.d}{4}$, and this is positive since
$(b+1)^{2} - d.d = b^{2} + 1 +2b - d.d$ can be expressed as
$2b^{2} + 2b - c.c - e.e$ (by using the fact that, $X^{T}X$
is symplectic and hence $b^{2} + d.d - c.c - e.e = 1$).
But $2b^{2} + 2b - c.c - e.e = b^{2} + d.d + 2b$ (since $b^{2} - c.c -
e.e = d.d$). But $b^{2} + d.d + 2b > 0$, since $b$, being one-fourth
the trace of the positive definite $X^{T}X$ is positive.
Hence both its roots are real. Further, since its
coefficients change sign ($\frac{b+1}{2}$ and $\frac{d.d}{16}$ are positive) 
both these roots are positive. 
Pick any one root and let $a$ be the positive square root of it.
This yields $a$ and hence $q$. 

To find $r$ and $p$, one inserts the expression $q = \frac{r\times p}{a}$
into the Equations for $c$ and $e$ and uses the vector triple product 
identities to find: 
\begin{eqnarray}
\label{firstlinsys}
c &=& 
2ap + \frac{2}{a}[(r.r)p - (p.r)r]\\ \nonumber
e &=& 
2ar + \frac{2}{a}[(p.p)r - (p.r)p]
\end{eqnarray}

This would yield a linear system for the unknows $p, r$ in terms of
the knowns $c$ and $e$ provided we can express $p.p. r.r, p.r$ in terms
of $e,c$. To achieve this, first note that
\begin{eqnarray*}
c.c &=& 4a^{2}p.p + 4\mid\mid q\times r\mid\mid^{2} + 8a^{2}p.(q\times r)
\\ \nonumber
e.e &=& 4a^{2}r.r + 4\mid\mid p\times q\mid\mid^{2} + 8a^{2}r.(p\times q)
\nonumber
\end{eqnarray*}.

Hence,
\[
c.c - e.e = 4a^{2}p.p - 4a^{2}r.r + 4\mid\mid q\times r\mid\mid^{2}
-  4\mid\mid p\times q\mid\mid^{2} 
\]

Next, using the identity $(u\times v). (\tilde{u}\times \tilde{v})
= (u.\tilde{u})(v.\tilde{v}) - (u.\tilde{v})(\tilde{u}.v)$, 
we find that
\[
\mid\mid q\times r\mid\mid^{2}
= (q.q)(r.r)
\]
(since $q.r = 0$)
and that
\[
\mid\mid p\times q\mid\mid^{2}
= (p.p)(q.q)
\]
(since $q.p = 0$)
 
Thus
\[
p.p - r.r = \frac{c.c - e.e}{4a^{2} - 4q.q}
\]
Since, we also know that $\frac{b+1}{2}
= a^{2} + q.q = 1 + p.p + r.r$, we have
that 
\[
p.p + r.r = \frac{b-1}{2}
\]
So, one gets a linear system for $p.p$ and $r.r$, in terms of
already determined quantities.
To find $p.r$, we compute that
\[
c.e = (2ap + 2q\times r).(2ar + 2p\times q)
= 4a^{2}(p.r) + 4(q\times r).(p\times q)
\]
Hence,
\[
c.e = 4a^{2}(p.r)
+ 4 [(q.p)(r.q) - (q.q)(p.r)]
= 4(a^{2}- q.q)(p.r)
\]
Now note that $a^{2} - q.q \neq 0$. Indeed, if it was zero, then
$a^{2} = \frac{d.d}{16a^{2}}$. But recall, the quadratic for
$a^{2}$ was $16a^{4} + d.d = 16 a^{2} \frac{b+1}{2}$. So 
if $a^{2} =  \frac{d.d}{16a^{2}}$, this would then imply that
$a^{2} = \frac{b+1}{4}$, which is not true.
So, we find $p.r = \frac{c.e}{ 4(a^{2}- q.q)}$.
This yields $p.r$.

Inserting these values for $p.p$, $p.r$ and $r.r$ into Equations
(\ref{firstlinsys}) yields a linear system for the vectors $p$ and $r$:

\begin{eqnarray}
\label{secondlinsys}
 c & = & \alpha p + \beta r\\ \nonumber
 e & = &  \beta p + \gamma r
\end{eqnarray}

with $\alpha = 2a + \frac{2r.r}{a} =  2a + \frac{b-1}{2a}
+ \frac{e.e - c.c}{a(4a^{2} - 4q.q)}$, $
\beta = -\frac{c.e}{4a^{2}-4q.q}$,
$\gamma =  2a +\frac{2p.p}{a}
= 2a + \frac{b-1}{2a} + \frac{c.c - e.e}{a(4a^{2} - 4q.q)}$,

The system Equation (\ref{secondlinsys}) is readily solved. 
Indeed, the system is invertible as the Cauchy-Schwarz
inequality reveals that the quantity    
$\alpha\gamma - \beta^{2}$ is at least $4a^{2}$.  

Thus, we have found $H$. If this $H$ is not positive definite, then
the $H$ corresponding to the other root of the Equation (\ref{quadratic})
has to be positive definite, since that is the only other $H$ which
is symplectic, symmetric, and with positive trace satisfying $H^{2}
= X^{T}X$. 

\vspace*{5mm}
 
\noindent {\it The case $d=0$} 
 Now Equation (\ref{quadratic}),
when $d=0$ has two roots, viz., $\frac{b+1}{4}$ (which is stricltly 
positive) and $0$. Thus, by picking $a= \frac{\sqrt{b+1}}{2}$, $q=0$,
$p = \frac{c}{\sqrt{b+1}}$ and $r = \frac{e}{\sqrt{b+1}}$, we find
the only $H$ which is symplectic, real symmetric, with positive trace
and which satisfies $H^{2} = X^{T}X$. Thus, by uniqueness, $H$ must be
the unique positive definite square root of $X^{T}X$.

In the constructive proof of the above theorem, a key step was an
explicit formula for the trace of $H^{2}$. This, calculation can be
used to obtain 
a complete characterization of positive definite symplectic $X$.
Furthermore, it indicates which root of Equation (\ref{quadratic})
to pick to compute the desired $H$.

\begin{theorem}
\label{positivedefinitesymplectic}
{\rm
Let $X = a(1\otimes 1) + p\otimes i + q\otimes j
+ r\otimes k$ be a real symmetric, symplectic matrix. Then it is
positive definite iff I) $a >0$, 
II) $2 a^{2} -2 q.q + 1 > 0$.
In particular, a symmetric and symplectic with matrix for which $a > 0, q=0$ is
positive definite.} 

\end{theorem}

\noindent {\it Proof:} 
It is well known that a real symmetric matrix is positive definite
iff all its eigenvalues are positive. Let, $P_{X}(x)
= x^{n} - a_{n-1}x^{n-1} + a_{n-2}x^{n-2} + \ldots 
+ (-1)^{n}a_{0}$ be the characteristic polynomial of $X$. Then,
by Descartes' rule of signs, for instance, $X$ is positive definite
iff $a_{i} > 0$ for all $i=0, \ldots , n-1$.
Since $X$ is symplectic, in addition, its characteristic polynomial
equals its reverse. Thus, we are guaranteed that $P_{X}(x)$
is of the form $P_{X}(x) = x^{4} - a_{3}x^{3} + a_{2}x^{2} - a_{3}x + 1$.
Now clearly $a_{3} = {\mbox Tr}(X) = 4a$.
It remains to find $a_{2}$. Now, $a_{2}
= \frac{1}{2}[({\mbox Tr}\  (X))^{2} - {\mbox Tr}\ (X^{2})]$.
But ${\mbox Tr}\ (X^{2})= 4(a^{2} + q.q + p.p + r.r)$. Using,
$1 = a^{2} + q.q - p.p - r.r$, we see that ${\mbox Tr}\ (X^{2}) = 
8a^{2} + 8q.q - 4$. This yields, 
\[
P_{X}(x) = x^{4} - 4ax^{3} + (4a^{2} - 4q.q + 2)x^{2} - 4ax + 1
\]
Hence positive definiteness of $X$ is equivalent to $a > 0$ and
$2a^{2} - 2q.q + 1 > 0$. In particular if $a > 0$ and $q=0$,
$X$ is positive definite.

From this follows a strengthening of Theorem (\ref{quaternionrepnI}), which is
worth stating separately:

\begin{theorem}
\label{quaternionrepnII}
{\rm Let $X\in Sp(4, R)$. Then there exist scalars $a, v_{0}, v_{2}$,
a unit quaternion $u$ and pure quaternions $p,q,r$
satisfying the
constraints $a > 0$, $2a^{2} - 2q.q + 1 > 0$,
$a^{2}-p.p + q.q - r.r = 1$, $q= \frac{r\times p}{a}$ 
and $v_{o}^{2} + v_{2}^{2}
= 1$ such that $X$ admits the following quaternion representation
$X = [u\otimes (v_{0} + v_{2}j)] 
[ a\ 1\otimes 1 + p\otimes i + q\otimes j +
r\otimes k]$.}
\end{theorem}

\begin{remark}
\label{squeezing}
Squeezing Transformations:
{\rm The authors of \cite{squeezing} argue persuasively that the
most general squeezing transformation is given by (the metaplectic 
representation of) a positive-definite symplectic matrix. Thus, the above
theorem obtains a quaternionic parametrization of such transformations for
two-mode systems.

More generally, any application which requires the non-compact part of
$Sp(4, R)$ would benefit from the parametrization in the above theorem,
\cite{squeezing,iwasawai,iwasawaii}. Note the matrices which are symplectic},
{\it but not orthogonal} {\rm [i.e., the non-compact portion of $Sp(4, R)$] 
do} {\it not} {\rm form a group. Hence, the polar decomposition is arguably
the most economical fashion to extract the non-compact part 
of an $X\in Sp(4, R)$.}  
\end{remark} 

\noindent From Theorem (\ref{quaternionrepnII}), 
it is now clear which root of Equation (\ref{quadratic})
to pick for ensuring a positive definite $H$, when $d\neq 0$. 
Clearly, the 
larger $a$ is, the smaller the $q.q$ is. Thus, the condition
$2a^{2}-2q.q + 1 > 0$ is more likely to be satisfied by the candidate
with the larger $a$. In fact, since there is precisely
only one positive definite square root, the $H$ corresponding to the
larger value of $a$ is positive-definite while the other candidate is not
(this can also be directly verified). So one should pick the larger value
for $a$, i.e.,  
one should pick $a$ to be the positive square root of $\frac{\frac{b+1}{2} +
\sqrt{\frac{(b+1)^{2}}{4} - \frac{d.d}{4}}}{2}$.

We summarize the above discussion into an algorithm for finding
the polar decomposition of an $X\in Sp(4,R)$:
\begin{algorithm}
\label{onlyalgorithm}
\vspace*{2mm} 
{\rm 
\begin{itemize}
\item 1. Compute directly $X^{T}X$ - this is just the Gram matrix
of the columns of $X$. $X^{T}X$ is positive definite (and thus symmetric)
and is also symplectic.
\item 2. Compute the quaternion representation of $X^{T}X$. This is
guaranteed to be of the form if $b(1\otimes 1) + c\otimes i + d\otimes j
+ e\otimes k$ (with $b\in R, c,d,e\in P$) since $X^{T}X$ is symmetric.
Furthermore, these quantities are linear in the entries of $X^{T}X$.
In addition, $b$, which is $\frac{1}{4}$ the trace of $X^{T}X$, is positive.

\item 3. Let $H$ be a matrix which has positive trace, real symmetric and
symplectic, and which satisfies $H^{2} = X^{T}X$. One such $H$ is the
positive definite factor in the polar decomposition of $X$. Let $H$, which
is symmetric, have quaternion representation $a(1\otimes 1) + p\otimes i + 
q\otimes j
+ r\otimes k$ (with $a\in R, p.q.r\in P$).    
Steps 4-6 below show how to compute $a, p, q, r$.

\item 4. If $d = 0$, then compute $a = \frac{\sqrt{b+1}}{2}$, $q=0$,
$p = \frac{c}{\sqrt{b+1}}$ and $r = \frac{e}{\sqrt{b+1}}$. This is guaranteed
to be the positive factor in the polar decomposition of $X$.

\item 5. If $d\neq 0$, then let $a$ be the positive square root of 
the larger of the two strictly positive roots of the quadratic
$x^{2} - \frac{b+1}{2}x + \frac{d.d}{16} = 0$. 
Define $q = \frac{d}{4a}$. Find $p, r$ by solving the linear system
of equations
Equation (\ref{secondlinsys}).
\item 6. This yields $H$ as the unique positive
definite square root of $X^{T}X$
and thus the symmetric part of the polar decomposition 
of $X$.

\item 7. Next compute $XH^{-1}$. For this no matrix inversion needs to
be performed. Instead use the fact that $H^{-1} = -J_{4}H^{T}J_{4}$.
This shows that if
\[
H = \left (\begin{array}{cc}
A & B\\
C & D
\end{array}
\right )
\]
is the block-form of $H$, then
\[
H^{-1} = \left (\begin{array}{cc}
D^{T} & -B^{T}\\
-C^{T} & A^{T} 
\end{array}
\right )
\]
Thus, $H^{-1}$ can be written by inspection.
The matrix $XH^{-1}$ is guaranteed to be symplectic and orthogonal, and thus
will admit the quaternion representation $u\otimes (v_{0} + v_{2}j)$,
with $u$ a unit quaternion and $v_{0}^{2} + v_{2}^{2} = 1$.

\item 8. To compute $u, v_{0}, v_{2}$ use the algorithm described
in \cite{FourportI}.

\end{itemize} 
}
\end{algorithm}

\vspace*{9mm}

\noindent
To give a different perspective on the ``paucity" of symplectic,
symmetric square roots of $X^{T}X$ when $d\neq 0$, a brief detour will be taken
through a calculation which is relevant for Theorem (\ref{charpolyII})
below as well. 

\noindent Thus, suppose that $H$ is a real symmetric, symplectic
matrix with non-zero trace satisfying $H^{2} = X^{T}X$, and suppose
$\tilde{H}$ is another. Then, since $H, \tilde{H}$ are square roots
of $X^{T}X$ [in the sense of Definition (\ref{sqrts})], there exists a real
orthogonal matrix $U$ with $\tilde{H} = HU$. Clearly $U$ must be symplectic
too. Further the condition, $\tilde{H}^{2} = H^{2}$ reads as   
$HUHU = HH$, which,
since $H$ is invertible,
is equivalent to \[
UH = HU^{T}\]
Similarly, the condition that $\tilde{H}$
be real symmetric is also equivalent to the same condition $UH = HU^{T}$.
Indeed, $(HU)^{T} = HU$ is equivalent to $U^{T}H = HU$, which
is equivalent to \[
H = UHU
\]
This in turn, is equivalent
to $UH = HU^{T}$.

\noindent So, considering $H$ as fixed we examine which symplectic orthogonal
$U$ lead to $UH = HU^{T}$. Let $U$ be represented as 
$u\otimes v$. Recalling that the imaginary part of $v$ is just
$v_{2}j$, we find that $UH = \alpha (1\otimes 1) + \beta\otimes p
+ \gamma\otimes j + \delta\otimes k + s\otimes 1 + 1\otimes t$, 
where the scalar $\alpha$ and the 
pure quaternions $\beta , \gamma , \delta , s, t$
are given by the equations
\begin{equation}
\label{expression1}
\alpha = au_{0}v_{o} + ({\mbox Im}\ u. q)v_{2}
\end{equation} 

\begin{equation}
\label{expression2}
\beta = u_{0}v_{0}p + v_{0} ({\mbox Im} \ u\times p)
+ u_{0}v_{2}r + v_{2} ({\mbox Im} \ u\times r)
\end{equation}

\begin{equation}
\label{expression3}
\gamma = u_{0}v_{0}q + v_{0}({\mbox Im} \ u\times q)
+av_{2} {\mbox Im} \ u
\end{equation}

\begin{equation}
\label{expression4}
\delta = -u_{0}v_{2}p - v_{2}({\mbox Im} \ u\times p)
+ u_{0}v_{0}r + v_{0} ({\mbox Im} \ u\times r)
\end{equation}

\begin{equation}
\label{expression5}
s = a ({\mbox Im} \ u)v_{0} - u_{0}v_{2}q -
v_{2} ({\mbox Im} \ u\times q)
\end{equation}

\begin{equation}
\label{expression6}
t = - [v_{0} ({\mbox Im}\ u).p + v_{2}({\mbox Im}\ u).r] {\bf i} 
+ [au_{0}v_{2} - v_{0}({\mbox Im}\ u).q]  {\bf j} 
+ [ v_{2}({\mbox Im}\ u).p - v_{0}({\mbox Im}\ u).r]  {\bf k} 
\end{equation}

A similar calculation for $HU^{T}$ reveals that the only
non-zero terms in $UH - HU^{T}$ are the following:
\begin{itemize}
\item An $1\otimes i$ term, viz.,
$-(2{\mbox Im}(u).p)v_{0} - (2{\mbox Im}(u).r)v_{2}$.

\item An $1\otimes j$ term, viz.,
$2au_{0}v_{2} - (2{\mbox Im}(u).q)v_{0}$.

\item An $1\otimes k$ term, viz.,
$2({\mbox Im}(u)).p v_{2} - 2 ({\mbox Im} (u).r)v_{0}$.

\item An $s\otimes 1$ term (with $s\in P$), viz.,
$2au_{0} {\mbox Im}(u) - 2u_{0}v_{2}q
-2v_{2}{\mbox Im}(u)\times q$.
\end{itemize}

Equating the first three terms to zero one finds that if $v_{0}\neq 0$,
then $\frac{v_{2}}{v_{0}} =
-\frac{{\mbox Im}(u).p}{{\mbox Im}(u).r} = 
\frac{{\mbox Im}(u).r}{{\mbox Im}(u).p} = 
\frac{{\mbox Im}(u).q}{a\alpha_{0}}$.

This forces ${\mbox Im}(u).r = {\mbox Im}(u). p = v_{2} = 0$ and
${\mbox Im}(u). q = 0$. Now as $H$ is symplectic, with its $q$ term
and its trace is not
zero, we have that $q$ is proportional to $r\times p$. So ${\mbox Im}(u)$,
being perpendicular to $p$ and $q$, must be parallel to $q$. But it is also
orthogonal to $q$. So ${\mbox Im}(u) = 0$. Hence $u$ is either
$+1$ or $-1$. Note that under these conditions the $s\otimes 1$ term also
vanishes. Finally,  $v_{2} = 0$ forces $v$ to be $+1$ or $-1$.
So, the only choices for $u\times v$, when $v_{0}\neq 0$, are
$1\otimes 1$ or $-1\otimes 1$.

If $v_{0} = 0$, then one finds ${\mbox Im}(u).p =
{\mbox Im}(u).r = 0$, again. Thus ${\mbox Im}(u)$ is proportional to $q$.
Furthermore, as $a\neq 0$ and $v_{2}\neq 0$, we get $u_{0} = 0$.
Thus, $u$ is purely imaginary and proportional to $q$. Once again,
the $s\otimes 1$ term vanishes uner these conditions. Thus, the only
choice for $u\times v$, in this case, are $\frac{q}{\mid\mid q\mid\mid}
\otimes j$ or $-\frac{q}{\mid\mid q\mid\mid}\otimes j$.

Thus, there are precisely four symplectic, real symmetric square roots
of $X^{T}X$ which have non-zero trace (two with positive trace, and
the remaining two with negative trace). If $H$ has positive trace, then
it is easy to see (e.g., by inspecting
the $1 \otimes 1$ term in $\tilde{H}$) that of the remaining three
candidates only the one corresponding to $U = \frac{q}{\mid\mid q\mid\mid}
\otimes j$  can also have positive trace.

\noindent {\it Calculating the Characteristic Polynomial of a Symplectic $X$}:
We now show how to calculate the characteristic polynomial of
a symplectic matrix, assuming its quaternion representation is available,
without any need for a determinant calculation. Of course, calculating
the quaternion representation is required. But since this provides the
polar decomposition also, it seems worthwhile to pursue this route for
the characteristic polynomial.

Thus, let $X$ be symplectic, with 
$[u\otimes (v_{0} + v_{2}j)] 
[ a\ 1\otimes 1 + p\otimes i + \frac{r\times p}{a}\otimes j +
r\otimes k]$, its quaternion representation. This now reads as the expression 
\[
\alpha (1\otimes 1) + \beta\otimes p
+ \gamma\otimes j + \delta\otimes k + s\otimes 1 + 1\otimes t\]
The scalar $\alpha$ and the
pure quaternions $\beta , \gamma , \delta , s, t$, have expicit expressions
given by Equation (\ref{expression1}), $\ldots$,
Equation (\ref{expression6}). 
 
Since its characteristic
polynomial $P_{X}(x)$ equals its reverse, it is of the form $P_{X}(x) =
x^{4} + a_{3}x^{3} + a_{2}x^{2} + a_{3}x +1$. Furthermore,
$a_{3} = - {\mbox Tr} (X)$ and $a_{2} = \frac{1}{2}
(({\mbox Tr}(X))^{2} - {\mbox Tr}(X^{2}))$.

Clearly, then 
\begin{equation}
\label{coefficient3}
a_{3} = -4\alpha  = -4(au_{0}v_{o} + ({\mbox Im}\ u. q)v_{2})
\end{equation} 

To find $a_{2}$ one needs the trace
of $X^{2}$. {\it A useful observation here is that for this one does
not need to calculate $X^{2}$.} Once $X$ has been found
the trace of $X^{2}$ is simply $4 (\alpha^{2} + \beta . \beta +
\gamma . \gamma + \delta . \delta - s.s - t.t)$.  

Now an explicit calculation (which makes repeated use of the fact that $u,v$
are unit quaternions) reveals
$\alpha^{2} + \beta . \beta +
\gamma . \gamma + \delta . \delta - s.s - t.t$ is the expression   
\[
(v_{0}^{2} - v_{2}^{2})[q.q + (u_{0}^{2} - \mid\mid {\mbox Im}\ u \mid\mid^{2})
a^{2}  - 2 ({\mbox Im}\ u.q)^{2}] + p.p + r.r - 2[({\mbox Im}\ u.p)^{2}
 + ({\mbox Im}\ u.r)^{2}] + 8au_{0}v_{0}v_{2} ({\mbox Im}\ u. q)
\]  

Hence the expression for $a_{2}$ is 
\begin{equation}
\label{coefficient2}
a_{2} = 8a^{2}u_{0}^{2}v_{0}^{2} + 8v_{2}^{2}({\mbox Im}\ u.q)^{2}
+ 2 (v_{2}^{2}- v_{0}^{2})
[q.q + (u_{0}^{2} - \mid\mid {\mbox Im}\ u \mid\mid^{2})
a^{2}  - 2 ({\mbox Im}\ u.q)^{2}] -2( p.p + r.r) + 4 [({\mbox Im}\ u.p)^{2}
 + ({\mbox Im}\ u.r)^{2}]   
\end{equation}

The above calculations can be summarized in
\begin{theorem}
\label{charpolyII}
{\rm Let $X\in Sp(4, R)$ be represented by quaternions as in Theorem
(\ref{quaternionrepnII}). Then its characteristic polynomial is
expressible as $P_{X}(x) = x^{4} + a_{3}x^{3} + a_{2}x^{2} + a_{3}x + 1$,
with $a_{3}$ and $a_{2}$ given by Equation (\ref{coefficient3}) and
Equation (\ref{coefficient2}) respectively.}
\end{theorem}

\begin{remark}
\label{diagonalize}
  
{\rm I) In computing the polar decomposition of a symplectic $X$,
one could resort to first diagonalizing the symmetric
matrix $X^{T}X$ in the form $V^{T} (X^{T}X) V = D$ 
(with $V$ orthogonal) and then finding
the unique positive definite square root, $Y$, in the form
$Y = VD^{\frac{1}{2}}V^{T}$. Here $D^{\frac{1}{2}}$ is the diagonal
matrix whose entries are the positive square roots of the entries
of the diagonal matrix $D$. For this, one could use the method of
\cite{ni}, which uses quaternion methods for diagonalizing
real symmetric $4\times 4$ matrices. Recall that the key 
step (though not the sole step)
in this is the following.
Given a real symmetric matrix $\hat{a}\ 1\otimes 1 +
\hat{p}\otimes i + \hat{q}\otimes j + \hat{r}\otimes k$, one has to
compute the singular value decompositon of the real $3\times 3$
matrix $Z = [\hat{p} \ \hat{q} \ \hat{r}]$, i.e, the matrix whose
columns are the vectors in $R^{3}$ equivalent to the pure quaternions
$\hat{p}, \hat{q}, \hat{r}$. In this regard, it is noted that this
calculation is somewhat simplified for the purpose at hand.   Indeed,
$X\in Sp(4, R)$ implies $X^{T}X\in Sp(4, R)$, i.e., the symmetric
matrix $X^{T}X$ is also symplectic, and thus the corresponding
terms $\hat{p}, \hat{q}, \hat{r}$ have the property that $\hat{q}$
is orthogonal to $\hat{p}, \hat{r}$. So when computing the SVD
of the corresponding $Z$, equivalently when forming the Grammian of
the triple of vectors  $\hat{p}, \hat{q}, \hat{r}$, one finds that
\[
Z^{T}Z = \left (\begin{array}{ccc}
\hat{p}.\hat{p} & 0 & \hat{p}.\hat{r}\\
0 & \hat{q}.\hat{q} &  0 \\
\hat{p}.\hat{r} & 0 & \hat{r}.\hat{r}
\end{array}\right )
\] 
Hence finding the SVD amounts to just diagonalizing
the $2\times 2$ matrix
\[
 \left (\begin{array}{cc}  \hat{p}.\hat{p} & \hat{p}.\hat{r}\\
\hat{p}.\hat{r} & \hat{r}.\hat{r} 
\end{array}
\right )\]
This can be done by hand. However, partly because, finding
the SVD of $Z$ is not the sole step, this approach was eschewed 
in favour of Algorithm (\ref{onlyalgorithm}).

\noindent II) Computing the Euler-Cartan Decomposition of $Sp(4, R)$:
The Euler-Cartan decomposition of $Sp(4, R)$ asserts that every
$X\in Sp(4, R)$ can be factorized as $U_{1}DU_{2}$, 
with $U_{i}$ symplectic and orthogonal, for $i=1,2$, and $D$ diagonal,
positive-definite and symplectic. This factorization can be explicitly
computed starting from the constructive polar decomposition given for
$Sp(4, R)$ in this work. Indeed, let $X=UP$ be the polar decomposition.
Following the technique in part I) of this remark, diagonalize the
positive definite symplectic matrix $P$. It is simple and interesting
to verify that the diagonalizing matrix $V$ yielded by this procedure
is actually symplectic and orthogonal. Thus, let $V^{T}PV = D$ be the
diagonalization of $P$, with $V$ symplectic and orthogonal, and $D$
positive definite symplectic. This yields $X = U_{1}DU_{2}$ with
$U_{1} = UV, U_{2} = V^{T}$. Clearly the 
$U_{i}$ are symplectic and orthogonal. This then yields (in closed form)
the Euler-Cartan Decomposition of $X$.
}
\end{remark}  
 
\section{Conclusions}
This work produced a quaternionic representation for the symplectic
group $Sp(4, R)$. Part of the utility of this work is that, in attempting
to provide explicit formulae for this representation, one obtains a very
simple and explicit technique for finding the polar decomposition of matrices
in $Sp(4, R)$. Since the positive definite factor in this decomposition
is one representation of the non-compact part of the symplectic group,
this circumstance can be used to address applications where this 
factor is relevant,\cite{squeezing,iwasawai,iwasawaii}.  
It would be an interesting exercise to 
derive other decompositions
of $Sp(4, R)$ starting from the representation provided here. 
   
Finding similar quaternionic representations of matrix groups preserving
other bilinear forms in dimension four is an interesting question.
It remains to see whether these lead to as elegant a set of expressions,
such as those for the symplectic group. Indeed, it is no
exaggeration to say that a pivotal role in the results here is played
by the fact that the $q$ term, in both Theorem (\ref{quaternionrepnI})
and Theorem (\ref{quaternionrepnII}), is essentially the cross product
of the $r$ and $p$ terms. Extending such representations to higher
dimensions is also open. Quaternionic representations are, of course,
limited to dimension four, just as there is no similar extension of
the pair of unit quaternions representation to higher dimensional
orthogonal groups. However, one can hope that in conjunction
with either numerical techniques for the symplectic group,
\cite{heike} or methods of Clifford Algebras, \cite{pertii} these
results can be extended to higher dimensions.


\begin{thebibliography}{99}

\bibitem{squeezing} Arvind, B. Dutta, N. Mukunda $\&$ R. Simon,
{\it Pramana}, {\bf 45}, 471 (1995).  

\bibitem{FourportI} T. Constantinescu, V. Ramakrishna, N. Spears, L. R. Hunt,
J. Tong, I. Panahi, G. Kannan, 
D. L. MacFarlane, G. Evans $\&$ M. P. Christensen,
{\it Journal of Optical Society of America A}, {\bf 23} (2006), 2919-2931.


\bibitem{goongi} G. Chen, D. Church, B. Englert, C. Henkel,
B. Rohnwedder, M. Scully $\&$ M. Zubairy,
{\it Quantum Computing Devices: Principles, Design and Analysis},
Chapman $\&$ Hall CRC Press, Boca Raton, (2006).

\bibitem{heike} H. Fassbender, {\it Symplectic Methods for the Symplectic
Eigenproblem}, Kluwer, New York (2000).
 
\bibitem{nii} H. Fassbender, D. Mackey $\&$ N. Mackey,
{\it Linear Algebra $\&$ its Applications}, {\bf 332}, 37, (2001).

\bibitem{gilmore} R. Gilmore, {\it Lie Groups, Lie Algebras and Some
of Their Application}, Dover Publications (2006).

\bibitem{iwasawai} I. Gjaja, {\it NEEDS 94}, World Scientific,
295 (1995).

\bibitem{iwasawaii} S. Habib $\&$ R. Ryne, {\it Phys. Rev Lett.},
{\bf 74}, 70 (1995).

\bibitem{haconi} D. Hacon, {\it SIAM J. Matrix Analysis },
{\bf 14}, 619, (1993).

\bibitem{hhorni} R. A. Horn $\&$ C. R. Johnson, {\it Matrix Aanlysis},
Cambridge University Press (1999).

\bibitem{hhornii} R. A. Horn $\&$  C. R. Johnson, 
{\it Topics in Matrix Aanlysis},
Cambridge University Press (1999).

\bibitem{kyf} C. R. Johnson, T. Laffe $\&$ C. K. Li, {\it Linear and Multilinear
Algebra}, {\bf 23}, 285, (1988). 

\bibitem{kuiper} J. Kuiper, {\it Quaternions and Rotation Sequences},
Princeton University Press (1999).

\bibitem{liebermann} P. Liebermann $\&$ C. Marle, {\it Symplectic
Geometry and Analytical Mechanics}, Reidel, Dordrecht (1987).

\bibitem{pertii} P. Lounesto, {\it Clifford Algebras and Spinors},
II edition, Cambridge University Press (2002).
  
\bibitem{ni} N. Mackey, {\it Siam J. Matrix  Analysis},
{\bf 16}, 421, (1995).

\bibitem{niii} D. Mackey, N. Mackey $\&$ S. Dunleavy,
{\it Electronic J of Linear Algebra}, {\bf 13}, 10, (2005).

\bibitem{structureddecompI} D. Mackey, N. Mackey $\&$ F. Tisseur,
{\it SIAM J. Matrix  Analysis},
{\bf 27}, 821, (2006).


\bibitem{expistruc} V. Ramakrishna $\&$ F. Costa, {\it J. Phys A:
Math $\&$ General}, {\bf 37}, 11613 (2004). 


\bibitem{expisufour} V. Ramakrishna $\&$  H. Zhou,
{\it J. Phys. A - Math $\&$ General}, {\bf 39}
3021, (2006).


\bibitem{iwasawaiii} T. Tam, {\it Applied Math. Lett.}, {\bf 19}, 1421 (2006). 

\end{thebibliography}
\end{document}